\documentstyle[psfig,epsfig,preprint,aps]{revtex}
\def\lsim{\raise0.3ex\hbox{$\;<$\kern-0.75em\raise-1.1ex\hbox{$\sim\;$}}}
\def\gsim{\raise0.3ex\hbox{$\;>$\kern-0.75em\raise-1.1ex\hbox{$\sim\;$}}}
\begin{document}
\draft
\title{ Neutrino Electron Scattering and Electroweak Gauge Structure: 
Future Tests}
\author{O. G. Miranda, 
\footnote{On leave from Departamento de F\'{\i}sica CINVESTAV-IPN.}
V. Semikoz
\footnote{On sabbatical leave from the Institute of the Terrestrial Magnetism,
the Ionosphere and Radio Wave Propagation of the Russian Academy of Science,
IZMIRAN, Troitsk, Moscow region, 142092, Russia. } 
and Jos\'e W. F. Valle
\footnote{E-mail valle@flamenco.ific.uv.es}}
\address{ Instituto de F\'{\i}sica Corpuscular - C.S.I.C.
Departament de F\'{\i}sica Te\`orica, Universitat de Val\`encia
46100 Burjassot, Val\`encia, SPAIN}

\maketitle

\begin{abstract}

Low-energy high-resolution neutrino-electron scattering experiments
may play an important role in testing the gauge structure of the
electroweak interaction. We propose the use of radioactive neutrino
sources (e.g. $^{51}$Cr) in underground experiments such as BOREXINO,
HELLAZ and LAMA. As an illustration, we display the sensitivity of
these detectors in testing the possible existence of extra neutral
gauge bosons, both in the framework of $E_6$ models and of models with
left-right symmetry.

\end{abstract}

\newpage
\section{Introduction}

Despite the success of the Standard Model (SM) in describing the
electroweak interaction, there have been considerable interest in
extensions of the gauge structure of the theory.  A lot of the
theoretical effort has been in models that can arise from an
underlying $E_6$ framework \cite{fae} as well as models with
left-right symmetry \cite{LR1,LR2}.

So far accelerator and reactor neutrino experiments have been the main
tool used in probing the gauge structure of the electroweak
interaction.  The first observation of $\nu_{e} e \to \nu_{e} e$
scattering at LAMPF \cite{Allen1} resulted in a total of $236 \pm 35$
events \cite{Allen2}.  The value for the total cross section was
$(10.0 \pm 1.5 \pm 0.9) \times 10^{-45} cm^2\times E_{\nu}(MeV)$ for a
neutrino mean energy of 30 MeV.  This measurement ruled out
constructive interference between neutral and charged currents, in
agreement with the Standard Model.  Moreover LEP measurements at the Z
peak have achieved very high precision in determining the neutral
current parameters. As for reactor experiments, electron anti-neutrino
fluxes are not accurate enough and have a bad geometrical factor
$\frac{\Delta \Omega}{4 \pi}=\frac12\frac{a^2}{r^2}$, where $a$ is the
size of the detector and $r\gg a$ is the distance from reactor.  As a
result reactors do not allow a precision test of the neutral current
weak interaction of the type we will describe here.

The $\nu_{e} e \to \nu_{e} e$ scattering process has proved to be an
useful tool in studying the $^8B$ neutrinos coming from the Sun at
underground installations.  The electron recoil spectrum has been
measured in the Kamiokande Cerenkov detector with a threshold energy
of 7.5 MeV.  Superkamiokande has reached a threshold energy of 6.5 MeV
and an energy resolution of about 20 \% .  Determining the electron
recoil spectrum should be also one of the goals to be pursued at the
future Sudbury Neutrino Observatory.

In this paper we focus on the possibility of studying $\nu_{e} e \to
\nu_{e} e$ scattering process from terrestrial neutrino sources with
improved statistics.  A similar idea has been suggested as a test of
non-standard neutrino electro-magnetic properties, such as magnetic
moments \cite{Vogel}.  In contrast to reactor experiments, a small
radioactive isotope source can be surrounded by gas or liquid
scintillator detectors with full geometrical coverage.  Here we
demonstrate that a low-energy high-resolution experiment can play an
important role in testing the structure of the neutral current weak
interaction.  The ingredients for doing such experiments are either
already available (e. g. the chromium source has already been used for
calibrating the GALLEX and SAGE experiments \cite{GALLEX}) or under
investigation (NaI detectors have already been used in dark matter
searches and the BOREXINO and HELLAZ detectors have been extensively
discussed).  These detectors should reach good energy resolution and
relatively low threshold.  Both BOREXINO and HELLAZ are planing
to detect neutrinos at energies below 1 MeV
\cite{Borexino,Hellaz}.  The BOREXINO solar neutrino detector should
have an energy threshold of 0.250 MeV and an estimated energy resolution
of about 12 \% at threshold \cite{Borexino}, while HELLAZ (Ref
\cite{Hellaz}) should have an energy threshold of 100 KeV and few \%
energy resolution. In the case of the LAMA proposal, they are planning
to use an ${}^{147}$Pm anti-neutrino source with a one tone NaI
detector in the energy region of 2-25 KeV.

We explicitly determine the sensitivity of these radioactive neutrino
source experiments as precision probes of the gauge structure of the
electroweak interaction and exemplify it in a class of $E_6$-type
models as well as models with left-right symmetry.

\section{The $\nu e$ scattering cross section}

We start our discussion of the neutrino electron scattering cross
section in a generic electroweak gauge model in which the main
contributions to this process arise from the exchange of charged and
neutral intermediate vector bosons, i.e. from charged (CC) and neutral
currents (NC).

The charged current amplitude for the $\nu_{e} e\to\nu_{e} e $  
process can be written, after a Fierz transformation as 
\begin{equation}
{\cal M} = \sqrt{2}G_{F} \bar{\nu}\gamma^{\mu}(1-\gamma_5)\nu 
	\bar{e}\gamma_{\mu} c_{L}\frac{1-\gamma_5}{2} e. \label{CCA}
\end{equation}
For the specific case of the SM we have $c_{L}=1$.

On the other hand, the neutral current contribution to the 
amplitude for the process $\nu_{e} e\to\nu_{e} e $ can be given as 
\begin{equation}
{\cal M} = \sqrt{2}G_{F} \bar{\nu}\gamma^{\mu}(1-\gamma_5)\nu
	 \bar{e}\gamma_{\mu}\big[ g_{L}\frac{1-\gamma_5}{2} + 
	g_{R} \frac{1+\gamma_5}{2}\big] e. \label{NCA}
\end{equation}
In this case the SM prediction is $g_{L,R}=\frac12 (g_{V}\pm g_{A})$,
$g_{V}=\rho_{\nu e}(-1/2+2\kappa \mbox{sin}^2\theta_{W})$ and
$g_{A}=-1/2\rho_{\nu e}$ where the $\rho_{\nu e}$ and the $\kappa$
parameters describe the radiative corrections for low-energy $\nu_{e}
e\to\nu_{e} e $ scattering, which have been recently computed by
Sirlin \cite{Sirlin} and taken into account in our calculations.  

The differential cross section for the process $\nu_e e\to \nu_e e$ in
terms of the effective amplitudes \ref{CCA} and \ref{NCA} is given by
\begin{equation}
\frac{d\sigma}{dT} = \frac{2m_{e}G^2_{F}}{\pi}\big\{ 
    (g_{L}+c_{L})^2+ g_{R}^2 - 
    [2g_{R}^2+\frac{m_e}{\omega_1} (g_{L}+c_{L})
    g_{R}]\frac{T}{\omega_1} + 
     g_{R}^2(\frac{T}{\omega_1})^2 \big\}  \label{DCS}
\end{equation}
Here, $T$ is the recoil electron energy, and $\omega_1$ is the
neutrino energy; therefore $T/\omega_1 <1$ for any value of
$\omega_1$. It is also important to note that we have in this
expression the ratio $m_e/ \omega_1$. Clearly all terms in this cross
section are potentially sensitive to corrections from new physics.
However, not all are equally sensitive to these corrections. In
particular, the linear $m_e/ \omega_1$ term will be important for low
energies and negligible for accelerator and reactor neutrino energies.
As we will show in section 4, testing for the presence of new physics 
should become feasible in the next generation of $\nu_{e}
e\to\nu_{e} e $ experiments with sufficiently low energies and high
resolution.

In the case of chiral \cite{331} contributions to the charged currents
due to new physics we can consider the value of $c_{L}$ to be the SM
prediction plus some new contribution
\begin{equation}
c_{L}=1+\delta c_{L}. 
\end{equation}
With this notation the differential cross section for this case can be
expressed, at first order in $\delta c_{L}$ as
\begin{eqnarray}
\frac{d\sigma}{dT} &\simeq& \frac{2m_{e}G^2_{F}}{\pi}\big\{ 
	(g_{L} +1 )^2+ g_{R} ^2 -
	(2g_{R} ^2 +\frac{m_{e}}{\omega_1} 
	(g_{L} +1 )g_{R})\frac{T}{\omega_1}+ 
	 g_{R} ^2 (\frac{T}{\omega_1})^2 \big\} + \\ \nonumber
	&+& \frac{2m_{e}G^2_{F}}{\pi}\big\{ 
	2(g_{L} +1 ) -
	\frac{m_{e}}{\omega_1} 
	g_{R}\frac{T}{\omega_1} \big\}\delta c_{L}  . \label{dif}
\end{eqnarray}

The first term in this equation corresponds exactly to the SM
expression while the next one is for the corrections due to new
physics. It is easy to see that in this case the effect of the new
contributions reduces to a shift in the value of the differential
cross section, while the shape is hardly affected, except for a linear
correction. In what follows we concentrate in neutral current
corrections that can arise in the framework of $E_6$ models and of
models with left-right symmetry.

\section{$E_6$ and  Left-Right Symmetric Weak Hamiltonian}

In this paper we consider $\nu_{e} e$ scattering as a test for
extensions of the Standard Model.  These extensions typically involve
an extra $U(1)$ symmetry at low-energies, as in the case of a class of
heterotic string inspired $E_6$ models \cite{fae}, or an extra
left-right symmetric $SU(2)$ \cite{LR1,LR2} at low-energies, such as
can arise in Grand Unified Theory (GUT) models such as SO(10).

In the case of models with an extra $U(1)$ hyper-charge symmetry may be
given an mixture of those associated with $U(1)_{\chi}$ and
$U(1)_{\psi}$, the symmetries that lie in SO(10)/SU(5) and
$E_6$/SO(10), respectively.  In table 1 we show the quantum numbers
for $Y_{\chi}$ and $Y_{\psi}$ for the SM particles.

The corresponding hyper-charge can be then specified by
\begin{equation}
Y_\beta =\mbox{cos}\beta Y_{\chi}+\mbox{sin} \beta Y_{\psi}, 
\end{equation}
while the charge operator is given as $ Q=T^3+ Y, $.  Any value of
$\beta$ is allowed, giving us a continuum spectrum of possible models
of the weak interaction. Here we focus on the most common choices
considered in the literature, namely $\mbox{cos} \beta =1$ ($\chi$ model),
$\mbox{cos} \beta =0$ ($\psi$ model) and $\mbox{cos} \beta =\sqrt{\frac38}$, $\mbox{sin}
\beta =-\sqrt{\frac58}$ ($\eta$ model). For definiteness we will 
also restrict ourselves to the case when only doublet and \mbox{sin}glet
Higgs bosons are present so that the tree level value of the $\rho$
parameter is one (these models were called {\sl constrained models} in
ref. \cite{npb345}). A more complete analysis over the whole range of
$\beta$ values, as well as the inclusion of the case when there is no
restriction on the Higgs sector can be carried out as in
Ref. \cite{Concha}. Moreover one might also consider the case when all
the Higgs bosons arise from the fundamental {\bf 27}-dimensional
representation of the primordial $E_6$ group (these models were called
{\sl constrained superstring models} in ref. \cite{npb345}). In the
latter case one would be able to determine the value of the $Z'$ mixing
angle in terms of the masses of $Z$ and $Z'$, thus leading to much
stronger constraints.

The amplitude for neutral $\nu e \to \nu e$ scattering in such models
is given by (see for example Ref. \cite{Concha})
\begin{equation}
{\cal M_{NC}} = \sqrt{2}G_{F} \bar{\nu}\gamma^{\mu}(1-\gamma_5) \nu
	\big\{ \bar{e}\gamma_{\mu}\big[ g_{L}\frac{1-\gamma_5}{2}+ 
	 g_{R} \frac{1+\gamma_5}{2}\big]\big\} e. 
\end{equation}
with
\begin{equation}
g_{L,R}=2\rho_{\nu e}[(v^{\nu e}_1(e) \mp 
a^{\nu e}_1(e))(v_1(\nu) - a_1(\nu))+ \gamma
(v^{\nu e}_2(e) \mp a^{\nu e}_2(e))(v_2(\nu) - a_2(\nu))] \label{concha}
\end{equation}
and 
\begin{equation}
\label{gammadef}
\gamma=\frac{M_{Z}^2}{M_{Z'}^2}
\end{equation}
and 
\begin{equation}
v^{\nu e}_1(e)=(-1/4+\kappa s^2_{W})\mbox{cos}\phi-s_{W}\frac{\mbox{cos}\beta}{\sqrt{6}}
\mbox{sin}\phi 
\end{equation}
\begin{equation}
v^{\nu e}_2(e)=(-1/4+\kappa s^2_{W})\mbox{sin}\phi+s_{W}\frac{\mbox{cos}\beta}{\sqrt{6}}
\mbox{cos}\phi 
\end{equation}
\begin{equation} 
a^{\nu e}_1(e)=\mbox{cos}\phi /4 +s_{W}(\frac{\mbox{cos}\beta}{\sqrt{24}}+
\sqrt{\frac58}\frac{\mbox{sin}\beta}{3})\mbox{sin}\phi 
\end{equation}
\begin{equation} 
a^{\nu e}_2(e)=\mbox{sin}\phi /4 -s_{W}(\frac{\mbox{cos}\beta}{\sqrt{24}}+
\sqrt{\frac58}\frac{\mbox{sin}\beta}{3})\mbox{cos}\phi 
\end{equation}
\begin{equation} 
v_1(\nu)=\mbox{cos}\phi /4-s_{W}(\frac{3\mbox{cos}\beta}{2\sqrt{24}}+
\sqrt{\frac58}\frac{\mbox{sin}\beta}{6})\mbox{sin}\phi  
\end{equation}
\begin{equation} 
v_2(\nu)=\mbox{sin}\phi /4+s_{W}(\frac{3\mbox{cos}\beta}{2\sqrt{24}}+
\sqrt{\frac58}\frac{\mbox{sin}\beta}{6})\mbox{cos}\phi  
\end{equation}
\begin{equation} 
a_1(\nu)=-\mbox{cos}\phi /4+s_{W}(\frac{3\mbox{cos}\beta}{2\sqrt{24}}+
\sqrt{\frac58}\frac{\mbox{sin}\beta}{6})\mbox{sin}\phi  
\end{equation}
\begin{equation}
a_2(\nu)=-\mbox{sin}\phi /4-s_{W}(\frac{3\mbox{cos}\beta}{2\sqrt{24}}+
\sqrt{\frac58}\frac{\mbox{sin}\beta}{6})\mbox{cos}\phi  
\end{equation}

We now turn to a brief discussion of the effective weak Hamiltonian
that arises in models based on the Left-Right Symmetric gauge group $$
G_{LR} \equiv SU (2)_L \otimes SU (2)_R \otimes U (1)_{B - L}.$$ These
models are theoretically attractive since they offer the possibility of
incorporating parity violation on the same footing as gauge symmetry
breaking, instead of by hand as in the Standard Model \cite{LR1,LR2}.

The coupling constants for neutral currents in the left-right
symmetric model (LRSM) are given by
\begin{equation}
g_{L,R}=A\frac12(g_{V}\pm g_{A})+B\frac12(g_{V}\mp g_{A}) \label{LRG}
\end{equation}
with 
\begin{equation}
A = \big( c_{\phi}-\frac{s^2_{W}}{r_{W}}s_{\phi}\big)^2 
	+\gamma\big( s_{\phi}+\frac{s^2_{W}}{r_{W}}c_{\phi} \big)^2
\end{equation}
\begin{equation}
B = \frac{c^2_{W}}{r_{W}}\big[ 
	-\big( c_{\phi}-\frac{s^2_{W}}{r_{W}}s_{\phi}\big) s_{\phi}+
 	\gamma\big( s_{\phi}+\frac{s^2_{W}}{r_{W}}c_{\phi} \big)\big] 
\end{equation}
where the shorthand notation $s_{\phi}=\mbox{sin}\phi$, $c_{\phi}=\mbox{cos}\phi$, 
$r_{W}=\sqrt{\mbox{cos}2\theta_{W}}$ has been used.

Note that $\nu_e e\to \nu_e e$ scattering is not sensitive to
right-handed charged currents because the interference term between
the corresponding amplitude and the Standard Model one is suppressed
either by the neutrino mass (Dirac case) or by the mixing with the
heavy neutrinos (Majorana case, seesaw model). In fact, this is just an
example of the general situation that one finds when trying to
constrain charged-current parameters via purely leptonic processes
(see refs. \cite{Holstein} and \cite{Wolfenstein}).

\section{Experimental Prospects}

The values of the coupling constants governing $\nu_e e\to \nu_e e$
scattering have been well measured from $e^{+}e^{-} \to l^{+}l^{-}$ at
high energies by the LEP Collaborations.  A combined fit from LEP
results at the Z peak gives \cite{Stickland} $g_{V}=-0.03805\pm
0.00059$ and $g_{A}=-0.50098 \pm 0.00033$.  These results have given
strong constraints on right-handed neutral currents, specially on
the mixing of the standard Z boson with other hypothetical neutral
gauge bosons, in the framework of global fits of the electroweak data
\cite{lang,lang1}.

In contrast we focus here on low-energy neutrino-electron scattering
experiments. These have been suggested in order to test unusual
neutrino electromagnetic properties, such as magnetic moments (see
\cite{Vogel}).  At present some experimentalists are considering the
possibilities of these kind of physics \cite{Fiorentini,Barabanov}.
Here we consider their role in testing the gauge structure of the
electroweak interaction.  Two things are required for this kind of
experiment: a strong low-energy electron-neutrino source and a
high-precision detector.

The first strong low-energy neutrino sources have been recently
prepared for the calibration of the GALLEX and SAGE \cite{GALLEX}
neutrino experiment. In the GALLEX case this was a $^{51}$Cr neutrino
source with an activity of $1.67 \pm 0.03$ MCi.  In Table 1 we show
the main characteristics of the ${}^{51}$Cr source.  Besides the four
different neutrino energy lines, there are also 320~KeV photons as
well as high energy (above 1~MeV) $\gamma's$ from impurities. The
GALLEX and SAGE collaborations addressed this problem using a tungsten
shielding in order to avoid radiological problems. This shielding
stopped not only the high energy but also the 320~KeV photons. The
size of the source, along with the shielding is close to one meter,
therefore we estimate that the neutrino flux just outside the
shielding is about  $\Phi=1.8 \times 10^{12} \nu/ cm^{2} sec$.
Other sources have also been proposed for calibration of low energy
neutrino detectors \cite{Haxton,Bahcall}.  For anti-neutrino sources,
among different possibilities, the $^{147}$Pm source \cite{Kornoukhov}
is the one that LAMA proposal is considering for its experiment on
neutrino magnetic moment searches.  This source has a maximum neutrino
energy of 234.7 KeV and it is planned to have an activity of 5-15
MCi. Its half-life is 2.6234 years. The expected size of the source
along with the shielding is 70cm. Thus one sees that the preparation
of high-quality isotope sources is not a problem, from our point of
view the main remaining problem seems to be the design of detectors
capable of making precise measurements using these sources.

At the moment no detector is able to measure the $\nu_{e} e \to
\nu_{e} e$ dispersion at energies below 1 MeV.  However, there are
several proposals in this direction. Here we will concentrate in
BOREXINO, HELLAZ and LAMA. Both BOREXINO and HELLAZ will be sensitive
to the required range of electron recoil energy for a ${}^{51}$Cr
source (The energy thresholds expected are 250 KeV and 100 KeV,
respectively) while LAMA will measure the $\bar\nu_{e} e \to
\bar\nu_{e} e$ dispersion for the $^{147}$Pm source.

In order to have a good determination of the differential cross
section we need to have small errors both in the recoil electron
energy and in the differential number of events. Hereafter we will
concentrate in the necessary resolution required in the experiments
in order  to be sensitive to the $Z'$ mass. 

The energy resolution of a detector depends on the energy range
itself, usually the higher the energy, the better resolution one
has. The reason for this is simple, if an electron has more energy
then it radiates more photons and then the uncertainties in the photon
counting decrease as $1/\sqrt{N_{photons}}$. This is valid for
photo-tubes and the same rule applies for other kinds of detectors.  A
characteristic quantity for the detector resolution is
$\Delta=\Delta_{\epsilon} \sqrt{T/\epsilon}$ with $\Delta_{\epsilon}$
being the energy resolution at the fixed energy $\epsilon$.
Experimentalists have estimated the resolution they can reach in the
three different proposals we are considering here. The corresponding
values of $\Delta_{1MeV} $ for the three different proposals are shown
in Table 3 along with other detector characteristics
\cite{Borexino,Hellaz,Barabanov}.

In the case of BOREXINO, bins of 50 KeV are envisaged by the
collaboration. The detector will be sensitive to the two main lines of
the ${}^{51}$Cr making up 90 \% of the neutrino flux; which is
equivalent to $\Phi=1.8 \times 10^{12} \nu/ cm^{2} sec$, if the source
is surrounded by the detector as was the case in the GALLEX
calibration experiment \footnote{We are assuming that the neutrino
source is placed at the centre of the detector; if this is not the
case, as in ref. \cite{Fiorentini}, the number of events would be
drastically reduced and therefore the statistical error will be too
large.}.

For HELLAZ the energy resolution is of the order of 3 \% and the
energy threshold is 100 KeV. This detector will consist of six tones
of helium ($N_{e}\simeq 2 \times 10^{30} $). Here we will assume that
the bin width will be 10 KeV.

Considering these parameters and a time period of twenty days, we need
an estimate of the expected differential number of events both for the
Standard Model and for the extended gauge models.  We have computed the
expectation for the differential number of events both in the Standard
Model as well as in extended models by using the expression
\cite{Bahcall}
\begin{equation}
\langle \frac{dN}{dT} \rangle =\Phi N_{e} \Delta t 
\int_{\omega_{min}(T_{trh})}^{\omega_{max}} f(\omega_1 )d\omega_1  
\int_{0}^{T_{max}(\omega_1 )} 
R(T,T')\frac{d\sigma}{dT}dT'
\end{equation}
with 
\begin{equation}
R(T,T')=\frac{MeV}{\sqrt{2\pi T'}\Delta_{1MeV}}
exp\left( \frac{-(T-T')^2}{2\Delta_{1MeV}^2T'}\right)
\end{equation}
In the absence of a direct measurement of the resolution function we
use a Gaussian, as advocated in \cite{Bahcall}.  

In the SM case we just need to substitute in the differential cross
section \ref{DCS} the SM expressions for $g_{L}$ and $g_{R}$, while for
the extended gauge theories we substitute the expressions given in
\ref{concha}, for the case of $E_6$ models, or \ref{LRG} for the
LRSM. For the case of a ${}^{51}$Cr source the energy spectrum
$f(\omega_1 )$ will be given by a sum of delta functions for the
different neutrino lines while for the anti-neutrino $^{147}$Pm we need
to consider the neutrino energy spectrum
\begin{equation}
f(\omega_1 ) d\omega_1 = \frac{1}{N}\omega_1^2 (W-\omega_1)
\sqrt{(W-\omega_1)^2-m_e^2)} d\omega_1
\end{equation}
with $N$ the normalization factor and $W$ equal to $m_e$ plus 234.7
KeV.  In this work we assume that the future experiments will measure
the Standard Model prediction, and we will make a fit on the extended
gauge model parameters by assuming that the measured number of events
per bin is given by
\begin{equation}
N_i=\int_{T_i}^{T_{i+1}} \langle\frac{dN}{dT}\rangle^{SM} dT
\end{equation}
In order to do such a hypothetical fit for the different models under
consideration we also need to know the total error per bin
$\sigma_i$. As we do not know this value we have considered different
values of $\sigma_i$ (in percent) in our analysis. We can only hope,
at this point, that they reflect in a realistic way both the
statistical and systematic errors. For a preliminary discussion of
backgrounds for Borexino see Fig 15 of their proposal \cite{Borexino}.
More precise measurements are now underway \cite{Shonert}.

As already mentioned, these experiments can not compete in sensitivity
with LEP results, therefore is not possible to improve the constraint
on the mixing angle with this kind of experiments, but they can play a
role in constraining the $Z'$ mass.  In our analysis we have fix the
mixing angle to be zero and we have fitted for the parameter $\gamma$,
under the assumptions specified above.

With the expressions given in \ref{concha} we can have the theoretical
prediction for the $E_6$ models, in particular for the $\chi$, $\psi$
and $\eta$ models. From \ref{LRG} we have the theoretical prediction
for the LRSM. With these theoretical predictions and with the
different values per bin we can make a fit on each model for different
hypothetical $\sigma_i$ values. The results, at 95 \% C. L. are shown
in Fig. 1 both for BOREXINO and HELLAZ. In this figure we also show
the constraint on $\gamma$ coming from a global electroweak fit
\cite{lang}.  One can see that the sensitivity is different
for different models, the most promising case being the $\chi $ model.
In this case an error of 5 \% in BOREXINO would provide a better
sensitivity than that obtained in a {\sl global} fit of the
electroweak data \cite{lang}, while for HELLAZ an error of 8 \% would
already give a better sensitivity than that of a {\sl global} fit.
For the case of $\eta$ model the situation is much less hopeful.
However, one can improve substantially the sensitivity by going to the
case of {\sl constrained superstring models}, since in this case the
$Z'$ mixing is determined by the $Z'$ mass \cite{npb345}. In the same 
figure we also have plotted the result for the HELLAZ experiment in
case they can cover the energy range from 100 KeV to 560 KeV instead
of the energy range of 100-260KeV. This most optimistic case
corresponds to the lower line in each one of the plots in Fig. 1.

One can see that HELLAZ could be more sensitive than BOREXINO in
testing new physics if they can control the systematic errors due to
their expected better energy resolution (larger number of
bins). However it is also important to notice that the expected
statistical error in HELLAZ will be bigger than in BOREXINO because of
the smaller detector mass. Therefore a good control of the systematic
errors is required.

In the BOREXINO case, its huge detector will give a small statistical
error, however, the energy resolution is poor, making it necessary to have a
good control on the systematic error in order to reach a meaningful
sensitivity to the $Z'$ mass, comparable to the present constraint
from a global electroweak fit.

We have repeated the same considerations made above for the case of
the recent LAMA proposal for the case of a 1 tone detector surrounding
a 10 MCi source with $4\pi$ geometry and 50 cm radius. Our results are
shown in Fig. 2. We have also shown by the black square dot the
constraint that one would get in the case when only the statistical
error is included for the specific configuration that we have
discussed here and one year running. One sees that the prospects for
getting a better constraint seem good if they can control systematic
errors.

\section{Discussion and Conclusions}

We have illustrated in this paper how the upcoming strong radioactive
neutrino sources and new low-energy detector technology are likely to
open a new window of opportunity for experimental searches which were
originally mainly directed to solar neutrino research.  In this work
we have considered the case of $\nu_{e} e \to \nu_{e} e$ scattering
and showed how these techniques could provide a better sensitivity
than achieved in present measurements.  This would allow stronger
tests of the electroweak interaction and, potentially, stronger
constraints on extended gauge theories.  We have considered three
particular detectors that could use isotope sources for studying this
process and we have found that there are good prospects for reaching a
better sensitivity on the $Z'$ mass than achievable at the moment, if
systematic errors can be put under control.

For example, for the LAMA proposal, assuming a one tone detector, a Pm
activity of 10 Mci and source of 50 cm radius we have determined the
number of events per 2 KeV bin which are expected in a year run as a
function of the recoil electron energy $T$ in the range 2-25 KeV. The
total number of events will be the sum over all bins. This is shown in
fig 3. Here we have also computed the Coulomb correction for the
anti-neutrino spectrum and found good agreement with the estimates
given in Fig. 1 in ref. [23]. Our calculations indicate that Coulomb
corrections may have an overall $\lsim$ 2\% effect in the number of
events shown in fig. 3, where these corrections have been included.
However, they do not affect our results in fig. 2, which are the
relevant ones for discriminating against new physics. In this figure
we have compared the relative sensitivity of LAMA for neutrino
magnetic moment searches \cite{Vogel} and extended gauge model (the
$\chi$ model, for definiteness) tests.  We have fixed, for
illustration, a neutrino magnetic moment of $\mu_{\nu} = 2.5\times
10^{-12}\mu_B$ and an extended neutral gauge boson mass of $M_{Z^{'}}=
330~GeV$. In contrast to the case of neutrino magnetic moments
(neglecting background) the sensitivity to the extended $Z'$ model
becomes competitive a little above 2 keV, and better for almost all
values of the recoil energy above this value. Even though the number
of events drops, one still has about 4\% sensitivity to new gauge
boson up to 40 KeV, where one expects one event per day. In other
words, for energies in the region $12~KeV\leq T\leq 25-30~KeV$ this
experiment should be sensitive to extensions of the standard model
neutral gauge structure, while the sensitivity to magnetic moment
becomes gradually lost.

How about improvements? In order to determine what improvements can be
achieved we have come back to the recoil energy range from 2-25 KeV
for which LAMA is optimum for neutrino magnetic searches. We have
studied the sensitivity of the LAMA experiment to a new neutral gauge
boson as a function of the {\sl total} error $\sigma$ that one can
reach. We have compared this sensitivity with the corresponding
sensitivity to a neutrino magnetic moment and have displayed in fig. 4
one sensitivity against the other.

Although this may seem far future and we do not wish to minimize the
experimental challenge posed by our proposal, one can see from this
figure that the sensitivity on new gauge boson increases much faster
than that on magnetic moment, because the first is linear in $\sigma$
whereas the second scales as $\sigma^{1/2}$.  In summary, we have seen
that future detectors will have the possibility of making stringent
low-energy tests of the electroweak interaction gauge structure
feasible.

\vspace{3cm} \noindent {\bf Acknowledgements} 

We would like to thank useful discussions with Igor Barabanov, Rita
Bernabei, Miguel Angel Garc\'{\i}a-Jare\~no, Concha
Gonz\'alez-Garc\'{\i}a, Philippe Gorodetzky, Francis von Feilitzsch,
Gianni Fiorentini, Vasili Kornoukhov, Sandra Malvezzi, Stanislav
Mikheev, Sergio Pastor, Stephan Schonert, and Tom Ypsilantis.  This
work was supported by DGICYT under grant number PB95-1077, by the TMR
network grant ERBFMRXCT960090 and by INTAS grant 96-0659 of the
European Union. O. G. M. was supported by a CONACYT fellowship from
the mexican government, and V. S. by the sabbatical grant SAB95-506
and RFFR grants 97-02-16501, 95-02-03724.

\newpage

\begin{table}
\caption{Quantum numbers for the light particles in the {\bf 27} of $E_6$.}
\begin{tabular}{cccc} \hline
  & $T_3$ &  $\sqrt{40}Y_{\chi}$ &  $\sqrt{24} Y_{\psi}$ \\
\hline
$Q$      & $\pmatrix{1/2 \cr -1/2 \cr}$  & -1   &  1  \\
$u^{c}$  & 0                             & -1   &  1  \\
$e^{c}$  & 0                             & -1   &  1  \\
$d^{c}$  & 0                             &  3   &  1  \\
$l$      & $\pmatrix{1/2 \cr -1/2 \cr}$  &  3   &  1  \\ 
\end{tabular}
\end{table}

\begin{table}
\caption{ Neutrino Energies and half-life for a ${}^{51}$Cr  source. }
\begin{tabular}{ccccc}
\hline
source & $\tau_{1/2}$ &  $\omega_1$ & & $T_{max}$ \\
\hline
${}^{51}$Cr & 27 days & 0.746 MeV & 81 \% & 0.559 MeV \\
& & 0.751 MeV & 9 \% & 0.563 MeV  \\
& & 0.426 MeV  & 9 \% & 0.268 MeV  \\
& & 0.431 MeV &  1 \% & 0.273 MeV   \\
\end{tabular}
\end{table}

\begin{table}
\caption{Characteritics of $\nu e$ Detectors 
\protect{\cite{Borexino,Hellaz,Barabanov}}}
\begin{tabular}{rrrr}
\hline 
 & $T_{th}$ (MeV)&$\Delta_{1MeV}$ & $N_{e}$  \\
\hline
BOREXINO  & $ .250$ & $\sqrt{1/300}$  MeV  &  $5\times 10^{31} $  \\
HELLAZ    & $ .1$   & $\sqrt{1/20000}$ MeV &  $2\times 10^{30} $  \\
LAMA      & $ .002$ & $.026$ MeV           &  $3\times 10^{29} $  \\
\end{tabular}
\end{table}

\newpage

\begin{figure}
\caption{Constraints on the the parameter $\gamma$ of 
eq. (9) as a function of the total error per bin that could be reached
in the BOREXINO (solid line) and HELLAZ experiments (the dotted line
is for the energy range from 100-260 KeV while the dashed one is for
100-560 KeV) for four different extended gauge models.  The
electroweak global fit constraint on $\gamma$ is also shown for
comparison.}
\label{neu}
\end{figure}

\begin{figure}
\caption{Constraints on the the parameter $\gamma$ of 
eq. (9) as a function of the total error per bin that could be reached
in the LAMA experiment for four different extended gauge models. The
black square dot shows the case when only the statistical error is
considered.  The electroweak global fit constraint on $\gamma$ is also
shown for comparison. }
\label{anti}
\end{figure}

\begin{figure}
\caption{Expected number of events in LAMA per electron recoil energy bin 
for the SM (square), for a neutrino magnetic moment of $\mu_{\nu} =
2.5\times 10^{-12}\mu_B$ (star), and for an extended gauge model
($\chi$ model) with a $Z'$ mass $M_{Z^{'}}=330~GeV$ (circle).}
\label{mugau2}
\end{figure}

\begin{figure}
\caption{Comparing the LAMA sensitivity to neutrino magnetic 
moment vs. $Z'$ mass in the $\chi$ model. }
\label{emu2}
\end{figure}

\newpage
\thispagestyle{empty}

\begin{figure}
\centerline{\protect\hbox{\psfig{file=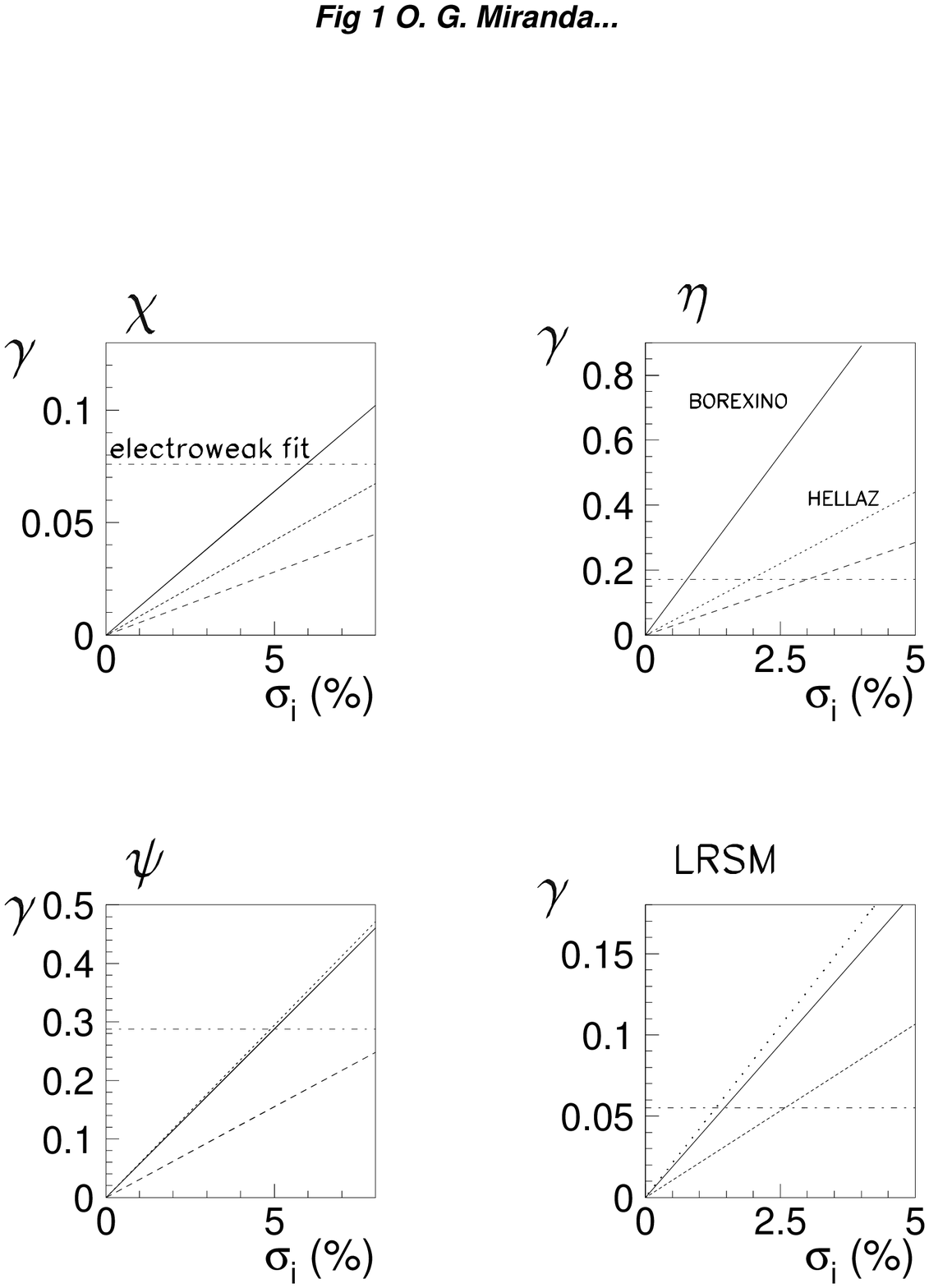,height=20.8cm,width=16cm}}}
\end{figure}

\newpage
\thispagestyle{empty}
\begin{figure}
\centerline{\protect\hbox{\psfig{file=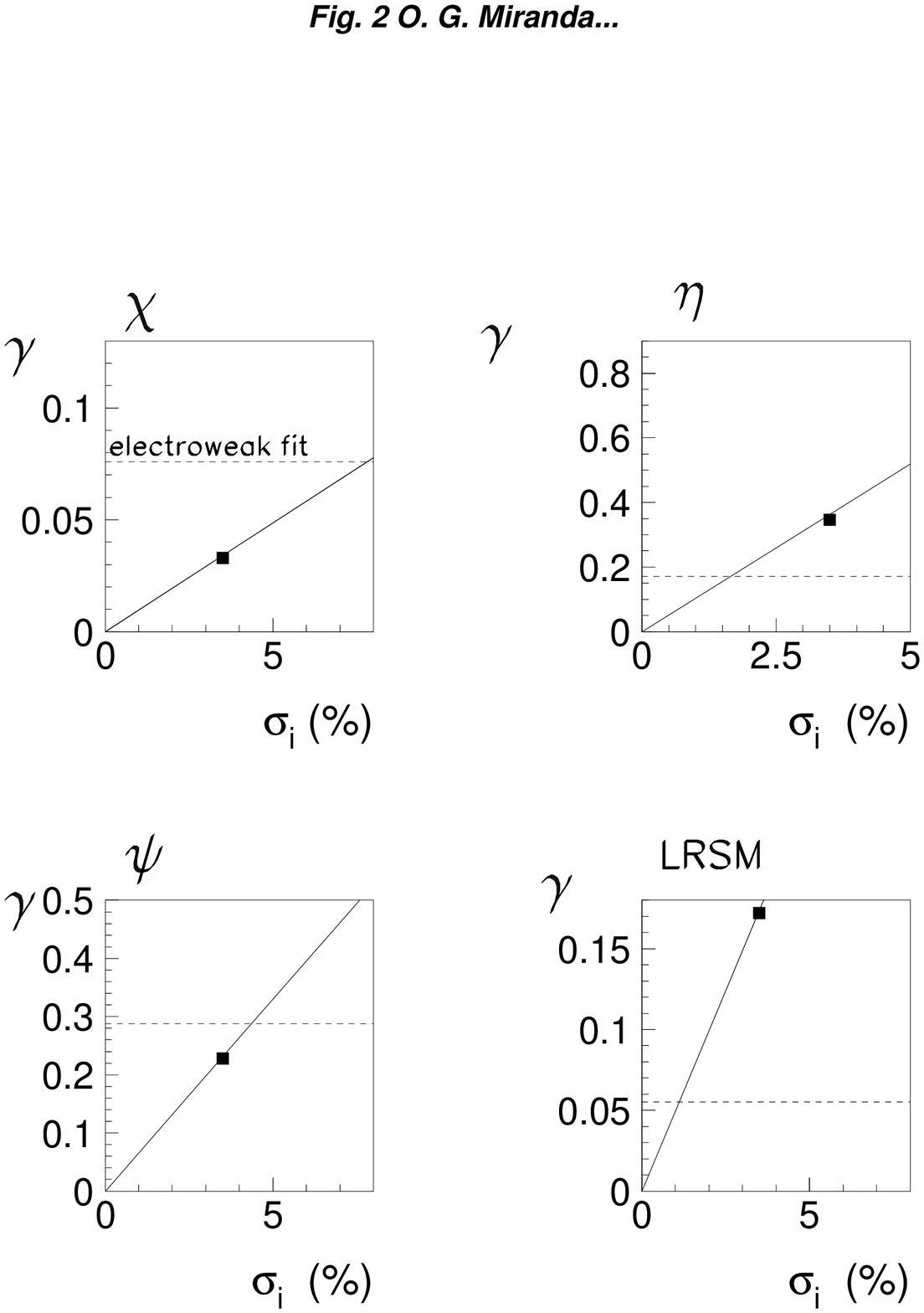,height=20.8cm,width=16cm}}}
\end{figure}

\newpage
\thispagestyle{empty}
\begin{figure}
\centerline{\protect\hbox{\psfig{file=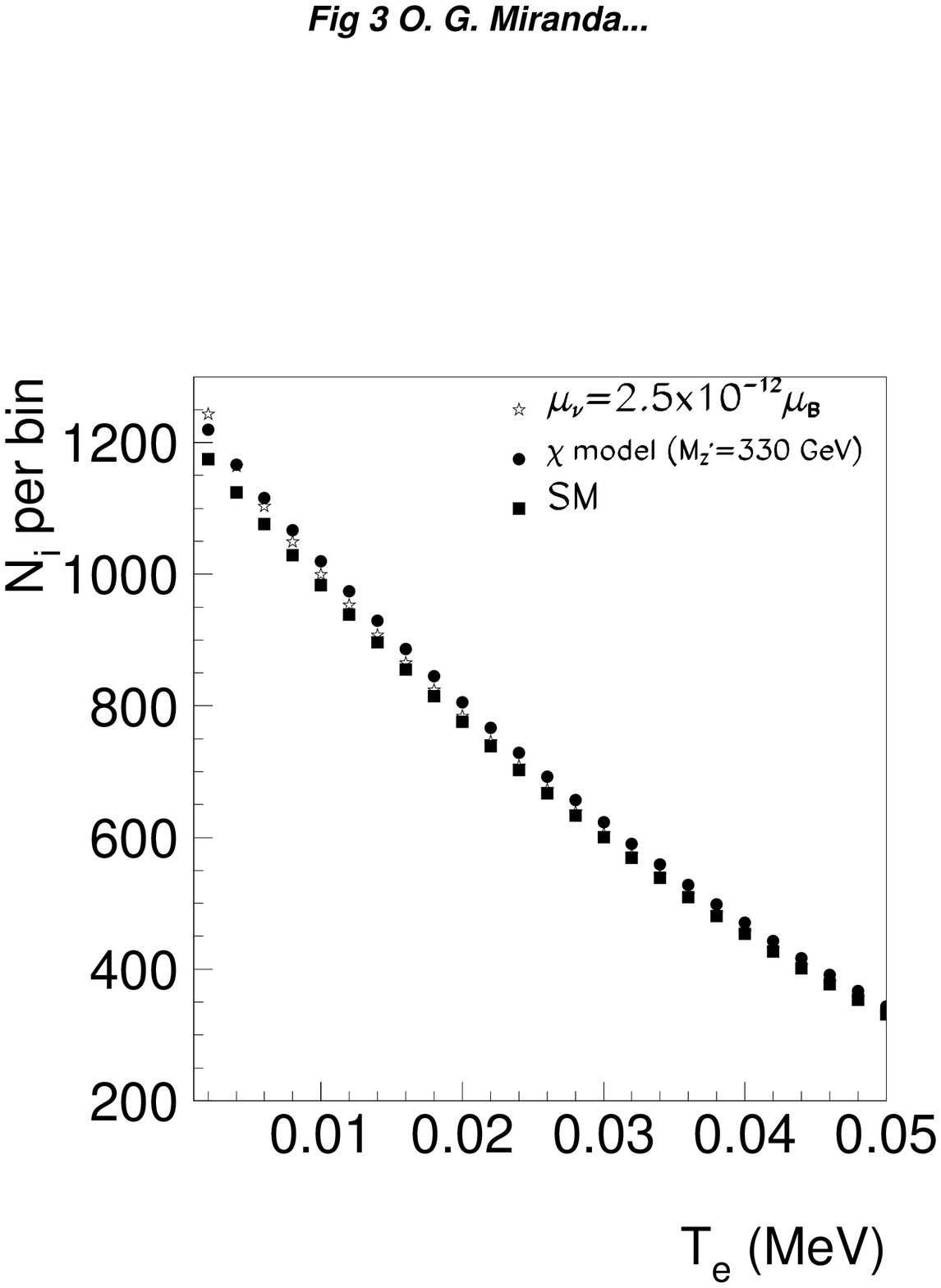,height=20.8cm,width=16cm}}}
\end{figure}

\newpage
\thispagestyle{empty}
\begin{figure}
\centerline{\protect\hbox{\psfig{file=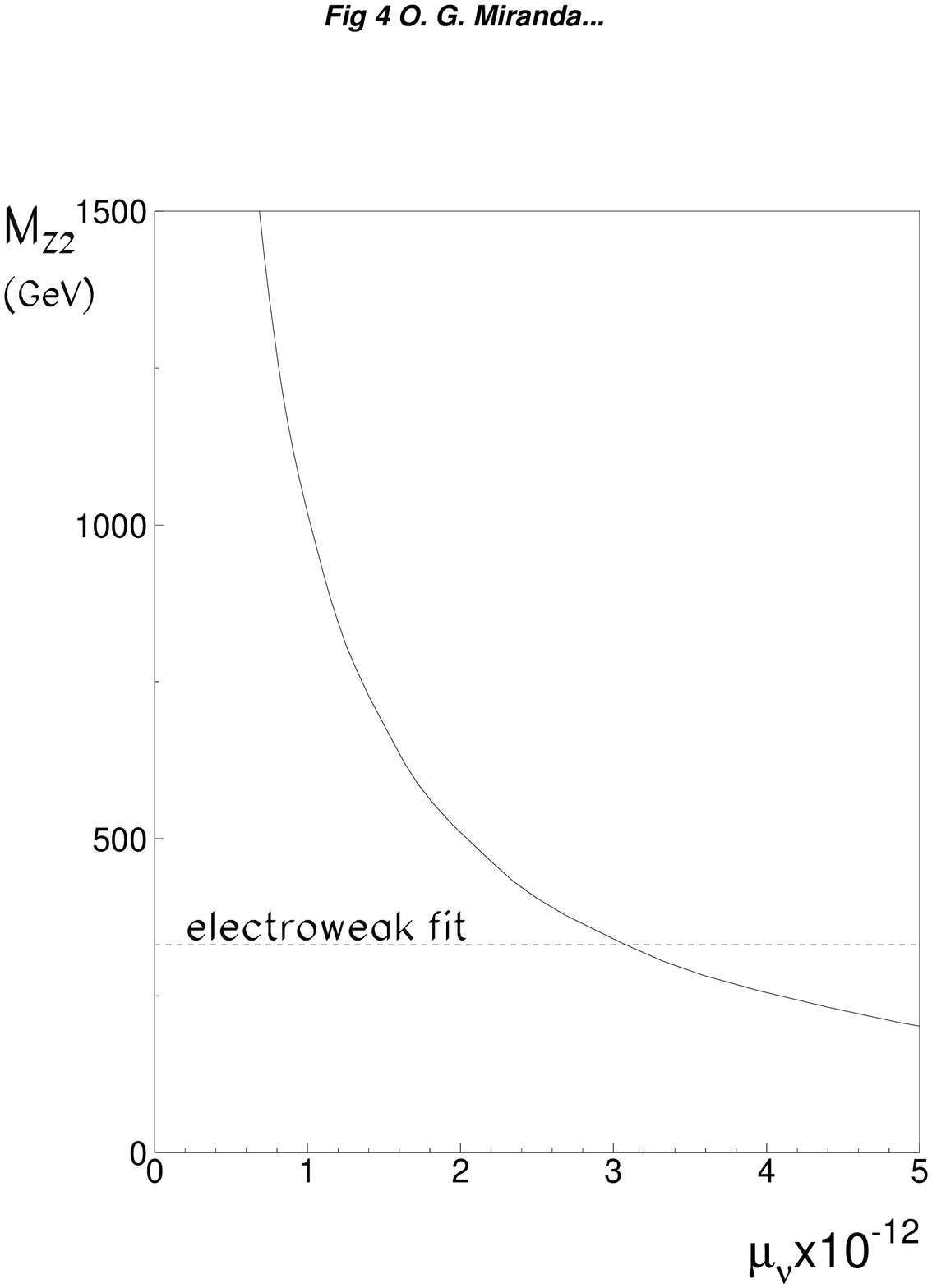,height=20.8cm,width=16cm}}}
\end{figure}

\end{document}